\documentclass[a4paper,11pt]{article}
\pdfoutput=1

\usepackage{jcappub}
\usepackage{amsfonts}
\usepackage{siunitx}

\newcommand{\dext}{\text{d}}
\newcommand{\der}{d}
\newcommand{\GN}{G}
\newcommand{\Mco}{M}
\newcommand{\Rco}{R_\text{m}}
\newcommand{\rhoNS}{\rho_\text{NS}}
\newcommand{\Rgrav}{R_\text{g}}
\newcommand{\Ecurv}{E_\text{curv}}
\newcommand{\dotEtot}{{\dot E}_\text{tot}}
\newcommand{\physrep}{Phys.\ Rep.}
\newcommand{\mnras}{Mon.\ Not.\ R.\ Astronom.\ Soc.}
\newcommand{\nat}{Nature}
\newcommand{\apj}{Astrophys.\ J.}
\newcommand{\sovast}{Soviet~Ast.}
\newcommand{\prd}{Phys.\ Rev.\ D}
\newcommand{\actaa}{Acta Astron.} 

\title{\boldmath An upper limit to the lifetime of stellar remnants from gravitational pair production}

 \author[a,1]{Heino Falcke,%
 \note{Jointly first authors.}}
 \author[a,b,1]{Michael F.\ Wondrak,}
 \author[b]{and Walter D.\ van Suijlekom}
 \affiliation[a]{Department of Astrophysics/IMAPP, Radboud University, \\
 P.O.\ Box 9010, 6500 GL Nijmegen, The Netherlands}
 \affiliation[b]{Department of Mathematics/IMAPP, Radboud University, \\
 P.O.\ Box 9010, 6500 GL Nijmegen, The Netherlands}
 
\emailAdd{h.falcke@astro.ru.nl}
\emailAdd{m.wondrak@astro.ru.nl}
\emailAdd{waltervs@math.ru.nl}

\abstract{
   Black holes are assumed to decay via Hawking radiation. Recently we found evidence that spacetime curvature alone without the need for an event horizon leads to black hole evaporation.
   Here we investigate the evaporation rate and decay time of a non-rotating star of constant density due to spacetime curvature-induced pair production and apply this to compact stellar remnants such as neutron stars and white dwarfs.
   We calculate the creation of virtual pairs of massless scalar particles in spherically symmetric asymptotically flat curved spacetimes. This calculation is based on covariant perturbation theory with the quantum field representing, e.g.,\ gravitons or photons.
   We find that in this picture the evaporation timescale, $\tau$, of massive objects scales with the average mass density, $\rho$, as $\tau\propto\rho^{-3/2}$. The maximum age of neutron stars, 
   $\tau\sim \SI{e68}{yr}$, 
   is comparable to that of low-mass stellar black holes. White dwarfs, supermassive black holes, and dark matter supercluster halos evaporate on longer, but also finite timescales.
   Neutron stars and white dwarfs decay similarly to black holes, ending in an explosive event when they become unstable. This sets a general upper limit for the lifetime of matter in the universe, which in general is much longer than the Hubble--Lema\^itre time, although primordial objects with densities above $\rho_\text{max} \approx \SI{3e53}{g/cm^3}$ should have dissolved by now. As a consequence, fossil stellar remnants from a previous universe could be present in our current universe only if the recurrence time of star forming universes is smaller than about $\sim \SI{e68}{years}$.}

\begin{document}
\maketitle
\flushbottom

\section{Introduction} 
Astronomy usually looks back in time when observing the universe, answering the question how the universe evolved to its present state. However, it is also a natural question to ask how the universe and its constituents will develop in the future, based on the currently known laws of nature. 

The evolution of stars is reasonably well understood and ends in the formation of either black holes, neutron stars, or white dwarfs. In binary systems, such compact remnants can evolve further via mass accretion, stripping, or mergers. Once they have settled into a single system, they are assumed to be stable and should not decay further on typical astronomical timescales. Adams \& Laughlin \cite{AdamsLaughlin1997a} investigated various scenarios for a future universe and derived a ``final stellar mass function,'' discussing possible long-term decay channels. 

It is generally believed that black holes will eventually decay via Hawking radiation \cite{Hawking1974a,Hawking1975a}, which can be illustrated by partial absorption and emission of virtual particle pairs created out of the quantum-field vacuum near the black hole event horizon. 

For neutron stars and white dwarfs, Reference~\cite{AdamsLaughlin1997a} considers dwindling via a hypothetical proton decay. At present we only have a lower limit to the proton lifetime \cite{NathFileviez-Perez2007a}. Reference~\cite{Caplan2020a} argues that in the absence of proton decay, the maximum lifetime of white dwarfs could be up to $\SI{e1100}{years}$. The main decay channel would be pycnonuclear fusion. 
Adams \& Laughlin \cite{AdamsLaughlin1997a} also mention the possibility of proton decay via virtual black holes and Zeldovich \cite{Zeldovich1976a} raises the possibility of gravitational annihilation of baryons, but both are not able to provide a proper timescale. This we attempt here.

Recently, we revisited Hawking radiation for the static Schwarzschild spacetime and found evidence that gravitational particle production by black holes does not require the presence of an event horizon~\cite{Wondrakvan-SuijlekomFalcke2023-PRL}, see also Refs.~\cite{MathurMehta2023a,Mathur:2024mvo}. 
The basic idea is that tidal forces in curved spacetimes can separate virtual particle pairs similar to the electric Schwinger effect where electric fields separate virtual charged pairs. This leads to emission of particles and a decay of the gravitational field together with its associated source. Indeed, such an intuitive explanation has already been invoked in standard textbooks (e.g., by  Shapiro \& Teukolsky \cite[pp.~367--368]{ShapiroTeukolsky1983a}, and  Birrell \& Davies \cite[p.~264]{BirrellDavies1984a}) and in articles~\cite{Dey:2017yez,Ong:2020hti}. 
Gravitational pair production also occurs in time-dependent spacetimes independent of an event horizon \cite{Parker:1968mv,ZeldovichStarobinskii1972a-updated,Visser:2001kq,HossenfelderSchwarzGreiner2003a,Vachaspati:2006ki}.

A further hint in this direction comes from a Gedankenexperiment involving Unruh radiation. According to Ref.~\citep{Unruh1976a}, the vacuum state of a quantum field is experienced as a thermal state by an accelerated observer. Hence, an accelerated observer, e.g. a flat plate, should see a thermal bath akin to Hawking radiation even if being an extended object~\cite{Lima:2018ifz}. If that observer, however, is the surface of a perfectly cold absorbing neutron star and Einstein's strong equivalence principle applies, then the surface should also see and absorb Unruh radiation. 
This would lead to a minimum temperature of neutron stars and, consequently, blackbody radiation. In the case of Unruh radiation for a macroscopic absorbing plate, the energy for this radiation would need to come from the acceleration of the plate and in the case of a neutron star surface from its gravitational acceleration, i.e., the neutron star's gravitational field. Hence, it is not immediately obvious that black holes would radiate and decay while neutron stars or white dwarfs would not.

The goal of this paper is therefore to gauge the astrophysical and cosmological implications of such a gravitational decay of compact objects which we idealize as constant-density objects. Indeed, using gravitational curvature radiation~\cite{Wondrakvan-SuijlekomFalcke2023-PRL}, we find that also neutron stars and white dwarfs decay in a finite time in the presence of gravitational pair production. We estimate their evaporation rates and spectra and discuss some implications.

\section{Gravitational particle production}
\label{sec:GravPtcleProd}
The production of particles can be inferred from the non-persistence of a quantum vacuum state \cite{Schwinger1951a}. 
This happens when more virtual pairs of particles and anti-particles are created than annihilated. One can calculate the probability that vacuum transitions to vacuum using the so-called 1-loop effective action which, in the Schwinger parametrization, corresponds to a Feynman path integral over all closed paths of virtual field excitations. 
If the resulting effective action has a vanishing imaginary part, all created pairs re-annihilate; if it contains a positive imaginary part, some pairs escape re-annihilation and become real particles. 
Our approach~\cite{Wondrakvan-SuijlekomFalcke2023-PRL} based on covariant perturbation theory~\cite{BarvinskyVilkovisky1990a} is capable of reproducing the electric Schwinger effect in the case of massless scalar electrodynamics, i.e., particle production in a purely electric background field. 
In the gravitational field of a black hole, the predicted production of massless particles comprises one component which resembles Hawking radiation with only one particle escaping, plus a new radiation component with both particles escaping. 

In a simple picture, both partners of a virtual pair explore different patches of a curved spacetime, their phases undergo a different time evolution, and hence they no longer completely annihilate when meeting again, leading to emission. This emission depends on the local spacetime curvature and happens at a distance compatible with the Heisenberg uncertainty relation which ensures overall causality and non-observability of negative energy. The energy of emitted particles is extremely low for macroscopic objects, implying a long timescale according to Heisenberg as we discuss in Sec.~\ref{sec:DiscussionConclusions} below.

The global event horizon in the case of a black hole spacetime only enters through determining the escape fraction of particles. 
For a particle escaping a Schwarzschild black hole which originated from between the event horizon at $2 G\Mco/c^2$ and the photon orbit at $3 G\Mco/c^2$ (at which the escape probability is exactly one half), the partner particle has to end up inside the black hole. Half a pair falling into the black hole and the other half escaping fits the popular picture often used to explain Hawking radiation. This process makes up roughly half of the total emission calculated by Ref.~\cite{Wondrakvan-SuijlekomFalcke2023-PRL} and also matches the level of emission predicted for Hawking radiation. For a particle escaping to infinity which originated from outside the photon orbit, also the entangled partner particle may escape the gravitational pull, which then differs from the Hawking picture.

Both emission components would essentially be unchanged if one replaced the black hole with a compact object of comparable size. 
If the surface radius $\Rco$ of such a putative non-rotating compact object of mass $\Mco$ and associated gravitational radius 
$\Rgrav \equiv G\Mco/c^2$ 
lay between the Buchdahl compactness limit~\cite{Buchdahl1959a} and the closed photon orbit radius, 
$9 \Rgrav/4 \leq \Rco \leq 3 \Rgrav$, 
then the first gravitational pair production component would still occur: For all pairs originating from between the surface and the photon orbit, at maximum one particle is emitted while the other one is absorbed. 
For all pairs originating from further out, or if the object had a radius $\Rco > 3 \Rgrav$ from the outset, then both particles of a creation event may escape. 

In the absence of an event horizon, there is pair production outside the object which leads to particles hitting the surface and also pair production inside the object. For our estimates, we assume those particles to be absorbed by the object and to increase and redistribute internal energy. Both components will lead to a surface emission, which is absent in black holes.

For our calculations, we use the sample case of a massless free real quantum scalar field.
The imaginary part of the 1-loop effective action in a curved spacetime reads, based on covariant perturbation theory including terms of up to second order in curvature\footnote{Taking into account terms to second order in curvature allows a proper description of particle production in static homogeneous electric fields and in the Schwarzschild geometry, but falls short of describing the Schwinger effect in the presence of magnetic fields, which would require higher-order terms and resummation~\cite{Wondrakvan-SuijlekomFalcke2023-Reply,FerreiroNavarro-SalasPla2023a-Comment}.} and truncated at an arbitrary order in the underlying proper time~$s$ \cite{BarvinskyVilkovisky1990a,CodelloZanusso2013a,El-Menoufi2016a},
\begin{align}
\begin{split}
\Im(W)
&= \frac{\hbar}{64\pi}\, \int \dext^4 x\, \sqrt{-g}\; \Bigg[ 
 \frac{1}{180} \left(R_{\mu\nu\rho\sigma} R^{\mu\nu\rho\sigma} 
  -R_{\mu\nu} R^{\mu\nu} \right)
+\frac{1}{2} \left(\xi -\frac{1}{6} \right)^2 R^2 
\Bigg].
\end{split}
\label{eq:Intro-EffAction}
\end{align}
Here $\xi$ is the gravitational coupling parameter, $g$ the metric determinant, and $R^{\,\mu}_{\,\;\nu\rho\sigma}$, 
$R_{\mu\nu} = R^{\,\alpha}_{\,\;\mu\alpha\nu}$, and $R$ are the Riemann curvature tensor, the Ricci tensor, and the Ricci scalar, respectively.%
\footnote{We follow the 
$(-,\,+,\,+,\,+)$ metric signature convention. An overdot corresponds to a time derivative with respect to the Schwarzschild-like time coordinate $t$ introduced below.}%
${}^\text{,}$%
\footnote{For the inclusion of the contact term $\square R$ beyond the derivation in~\cite{Wondrakvan-SuijlekomFalcke2023-PRL}, with $\square = \nabla_\mu \nabla^\mu$ being the Laplacian, see the discussion in \cite{Chernodub2023a}. 
Using the Ricci decomposition, one can show that the effective action only comprises positive coefficients for 
$\xi \in [0,\,(1-\sqrt{2/15})/6]$ evaluated on a metric of form~\eqref{eq:MetricAnsatz}. 
Furthermore, it stays positive over the whole parameter range of interest,
$\xi \in [0,\,1/6]$,
for the interior and exterior Schwarzschild solution.}

In this paper, we investigate simplified optically thick, non-rotating, spherically symmetric compact objects of mass $\Mco$ and radius $\Rco$ embedded in an asymptotically flat classical vacuum spacetime. In particular, we consider the case of an incompressible fluid, i.e., with a constant density 
$\rho = \Mco/(4\pi/3) \Rco^3$, referred to as interior Schwarzschild solution~\cite{Schwarzschild1916b-adapted}. 
The associated line element in Schwarzschild-like coordinates ($t$, $r$, $\theta$, $\phi$) arises from the Tolman--Oppenheimer--Volkoff equation \cite{Tolman1939a,OppenheimerVolkoff1939a} as 
\begin{equation}
\label{eq:MetricAnsatz}
\der s^2
= g_{tt}\, \dext (ct)^2 +\left(1 -\frac{2\GN m(r)}{c^2\, r}\right)^{-1}\! \dext r^2 
  +r^2 \left( \dext \theta^2 +\sin^2 \theta\, \dext \phi^2 \right)
\end{equation}
with the mass function
$m(r)
= \int_0^r\! \dext \bar{r}\, 4\pi \bar{r}^2\, \rho
= \Mco r^3/\Rco^3$.
The metric component $g_{tt}$ reads in the interior ($r \leq \Rco$)
\begin{align}
g_{tt}
&= -\frac{1}{4}\, {\left(
  3\sqrt{1 -\frac{2\Rgrav}{\Rco}} -\sqrt{1 -\frac{2 \Rgrav r^2}{\Rco^3}}
 \right)}^2
\end{align}
and $g_{tt} = -(1- 2\Rgrav/r)$ in the exterior.
The spacetime geometry is uniquely characterized by the parameters $\Mco$ and $\Rco$. The ratio thereof is referred to as compactness parameter 
$C \equiv \GN \Mco/c^2 \Rco$. Its inverse, $C^{-1}=\Rco/\Rgrav$, presents the object size in units of the gravitational radius. 
Such a matter configuration can satisfy Buchdahl's bound on the highest possible compactness for classical objects, 
$C \leq 4/9$~\cite{Buchdahl1959a}. Note the gap to the compactness of black holes, e.g., $C = 1/2$ for Schwarzschild (see also the review in Ref.~\cite{CardosoPani2019a}).

\begin{figure}[!ht]
\centering
\includegraphics[width=.5\linewidth]{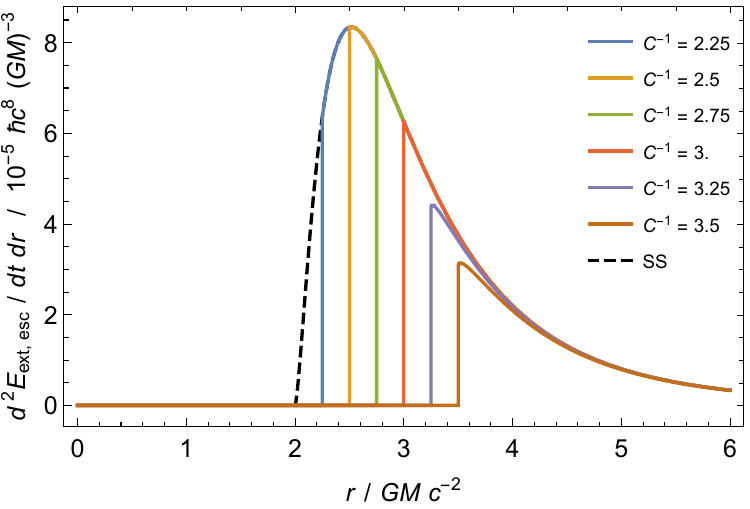}
\caption{\label{fig:dEextEscBydtdrVsrCompSS} Radial source distribution of energy flux directly escaping to infinity, 
$\der^2 E_\text{ext,esc} / \der t\, \der r$. The corresponding Schwarzschild distribution is shown as reference (black dashed).}
\end{figure}

The rate density of particle production events,
$2 \Im (\mathcal{L}_\text{eff})/\hbar$,
directly follows from the effective Lagrangian 
$\mathcal{L}_\text{eff}$
which is the integrand of the effective action $W$. It corresponds to the rate density of produced particles upon correcting for the number of pairs created per event, which can be estimated as
$f_\text{ppe} = 12/\pi^2$ 
in analogy to the Schwinger effect.
When focusing on the exterior emission, only those massless real particles escape the gravitational pull whose initial direction lies outside the absorption cone of opening angle $\alpha_\text{capt}$, providing an additional correction factor 
$f_\text{esc} = (1 +\cos \alpha_\text{capt})/2$ for intrinsically isotropic emission.
For $\Rco \leq 3\Rgrav$, i.e., if the compact object fits within its closed photon orbit, the expression for the opening angle agrees with the Schwarzschild case~\cite{Synge1966a}. 
For $\Rco > 3\Rgrav$ we derive
$\sin^2 \alpha_\text{capt}
= \Rco^3/r^3 \times (r -2\Rgrav)/(\Rco -2\Rgrav)$.
Finally, for the Ricci-flat case the characteristic energy per emitted particle for a local static observer can be estimated to be
$\Ecurv 
= \hbar c\, \left(K(r)/{30}\right)^{1/4}$  \cite{Wondrakvan-SuijlekomFalcke2023-PRL}
using the Kretschmann scalar
$K = R_{\mu\nu\rho\sigma} R^{\mu\nu\rho\sigma}$,
which is a measure of local spacetime curvature. In the exterior, it reads
$\Ecurv 
= 2^{3/4} \hbar c/5^{1/4}\Rgrav \times (r/\Rgrav)^{-3/2}$.
Particles produced in the exterior spacetime lead to an energy flux (see \cite{Wondrakvan-SuijlekomFalcke2023-PRL}) given by
\begin{align}
\label{eq:Formula_E_flux_ext}
\frac{\der E_\text{ext}}{\der t}
&= \frac{16\pi c}{\hbar}\, \int_{\Rco}^\infty \dext r\; 
  r^2\, f_\text{ppe}\, \frac{\Ecurv}{\sqrt{-g_{tt}}}\, \Im (\mathcal{L}_\text{eff}).
\end{align}
This includes the component ``ext,esc'' escaping to infinity, which is obtained when additionally considering the factor $f_\text{esc}(r)$ in the integral. 
Figure~\ref{fig:dEextEscBydtdrVsrCompSS} shows the radial emission profile for the energy flux escaping to infinity due to pair production outside the stellar remnant.

For estimating the characteristics of the emission, we assume that the compact object is optically thick absorbing all the particles produced in the exterior which cannot escape, ``ext,abs'', and absorbing those produced in the interior, ``int.'' The latter component has the energy flux
\begin{align}
\label{eq:Formula_E_flux_int}
\frac{\der E_\text{int}}{\der t}
&= \frac{16\pi c}{\hbar}\, \int_0^{\Rco} \dext r\; 
  r^2\, f_\text{ppe}\, \frac{\Ecurv}{\sqrt{-g_{tt}}}\, \Im (\mathcal{L}_\text{eff}).
\end{align}
The components $\der E_\text{ext,abs}/\der t$ and $\der E_\text{int}/\der t$ increase and redistribute the internal energy of the compact object and, thermodynamic equilibrium provided, lead to a surface emission with energy flux
$\der E_\text{surf}/\der t = \der E_\text{ext,abs}/\der t + \der E_\text{int}/\der t$
possessing a blackbody spectrum for a local Schwarzschild observer.
In contrast, the radiation component mentioned first, $\der E_\text{ext,esc}/\der t$, corresponds to direct emission similar to the black-hole case. 
The total energy flux is then given by
$\der E_\text{tot}/\der t
= \der E_\text{ext}/\der t +\der E_\text{int}/\der t
= \der E_\text{ext,esc}/\der t +\der E_\text{surf}/\der t$.

\begin{figure}[th]
\centering
\includegraphics[width=.5\linewidth]{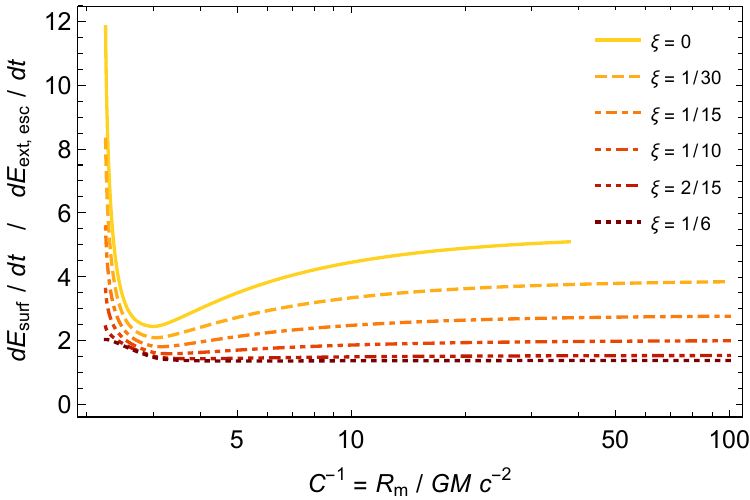}
\caption{\label{fig:PowerRatioSurfToExtEscVsInvCompactness} Ratio between the surface energy flux from a spherical mass of constant density to the escaping external energy flux as a function of the radius of the compact object in gravitational radii (for the allowed region $C^{-1} \geq 9/4$) and for different values of the gravitational coupling parameter $\xi$.
}
\end{figure}

Figure~\ref{fig:PowerRatioSurfToExtEscVsInvCompactness} compares the surface energy flux, $\der E_\text{surf}/\der t$, with the energy flux from direct emission, $\der E_\text{ext,esc}/\der t$, as a function of the inverse compactness for different values of the coupling parameter~$\xi$. 
For emission of graviton-like particles (minimally coupled, $\xi=0$), the interior contribution can be an order of magnitude higher, while for photon-like emission (conformally coupled, $\xi=1/6$) it can be higher by a factor $\sim 1.5$. 
In this graph, two effects are relevant: On the one hand, for small values of the inverse compactness, a larger fraction of externally produced particles gets absorbed by the compact object and is re-radiated via surface emission. On the other hand, for large values of the inverse compactness, the curvature in the exterior region of spacetime is markedly reduced so that most particle emission happens in the interior. This applies in particular to the situation of smaller coupling parameters $\xi$ which are more sensitive to volume changes by curvature.

\section{Results}
As an example, we will now use Eqs.~\eqref{eq:Formula_E_flux_ext} \& \eqref{eq:Formula_E_flux_int} to calculate the emission for an opaque object of constant density, e.g., a toy model for a neutron star, keeping the mass $M$ and surface radius $\Rco$ as free parameters. 
Given that the wavelength of gravitational curvature radiation typically is of the order of the spacetime curvature radius and stellar radius, one might expect a lower optical depth in a detailed model for thermal transport in neutron stars. Our main focus, however, is on order-of-magnitude estimates. 

\begin{figure}[!ht]
\centering
\includegraphics[width=.5\linewidth]{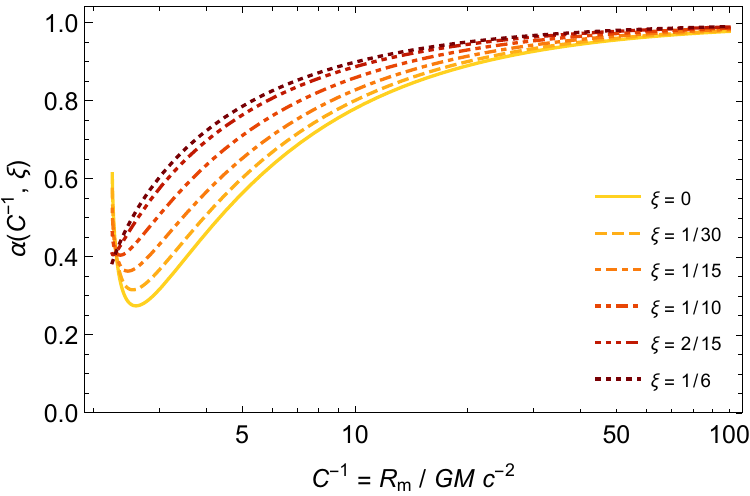}
\caption{\label{fig:AlphaNumCorrFctVsInvCompactness} Numerical correction function $\alpha$ for the total emitted power of a compact object, as introduced in Eq.~\eqref{eq:EnergyfluxTotal}, which takes into account redshift and absorption by the central object.}
\end{figure}

Extracting the main dependency of Eqs.~\eqref{eq:Formula_E_flux_ext} \& \eqref{eq:Formula_E_flux_int} with mass and radius, we can write the total energy flux scaled to typical parameters of a neutron star as
\begin{align}
\label{eq:EnergyfluxTotal}
\dotEtot
&= \SI{2.4E-22}{erg/s}\; \frac{f(\xi)}{0.38} \; \alpha\!\left(\frac{\Rco}{\Rgrav},\xi\right)\, \left(\frac{M}{1.44\, M_{\odot}}\right)^{-2}\left(\frac{\Rco}{6\,\Rgrav}\right)^{-9/2}\\
&= \SI{2.4E-22}{erg/s}\; \frac{f(\xi)}{0.38} \; \alpha\!\left(\frac{\Rco}{\Rgrav},\xi\right)\, \frac{M}{1.44\, M_{\odot}}\, \left(\frac{\rho}{\rhoNS}\right)^{3/2}
\end{align}
with $\rhoNS = \SI{3.3e14}{g/cm^3}$ and 
the asymptotic ($\Rco/\Rgrav \to \infty$) scaling function
\begin{equation}
\label{eq:f_xi}
f(\xi)
= 1 -\frac{2700}{285 +2^{9/2}\, 5^{3/4}}\, \xi +\frac{8100}{285 +2^{9/2}\, 5^{3/4}}\, \xi^2,
\end{equation}
where $f(1/6)\approx 0.38$.
The numerically calculated correction function $\alpha(C^{-1},\xi)$ is presented in Fig.~\ref{fig:AlphaNumCorrFctVsInvCompactness}. It is almost independent of the coupling parameter $\xi$, converges to unity for $C^{-1} \to \infty$, and stays of order unity for all allowed values of compactness. 

For comparison, a black hole with a mass of $1.44\, M_{\odot}$ has 
$\dot E_\text{tot}=\SI{2.7e-21}{erg/s}$, 
scaling with $M^{-2}$. This is a factor 7--14 more than for a neutron star with $\Rco = 6\,\Rgrav$. 

\begin{figure}[!ht]
\centering
\includegraphics[width=.5\linewidth]{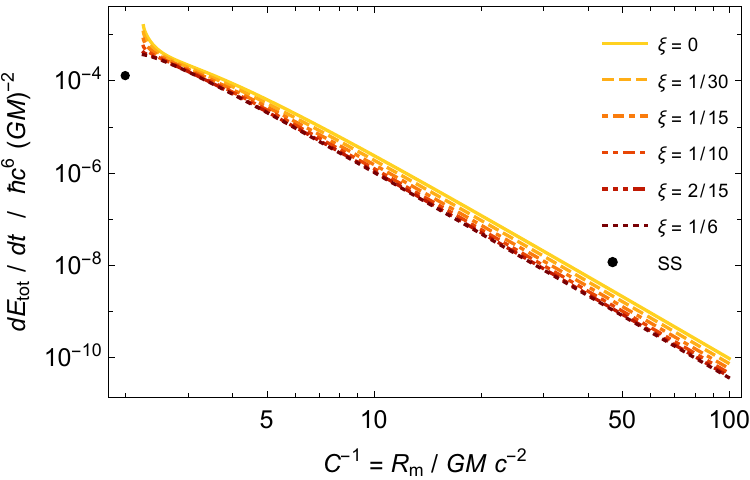}
\caption{\label{fig:PowerTotalVsInvCompactness} Total energy flux emitted by a spherically symmetric optically thick object of constant density as a function of inverse compactness (for a neutron star, $C^{-1} \sim 6$).
The black dot at $C^{-1} = 2$ represents the Schwarzschild reference value.}
\end{figure}

Figure~\ref{fig:PowerTotalVsInvCompactness} displays the total energy flux (direct and surface emission) produced by objects of different values for the inverse compactness, starting from the Buchdahl limit, $C^{-1} = 9/4$. The coupling parameter $\xi$ leads to a scaling by a factor of at maximum $2.5$, in accordance to Eq.~\eqref{eq:f_xi}. For large values of the inverse compactness, the energy flux scales as $(C^{-1})^{-9/2}$ in agreement with Eq.~\eqref{eq:EnergyfluxTotal}.

\begin{figure}[th]
\centering
\includegraphics[width=.5\linewidth]{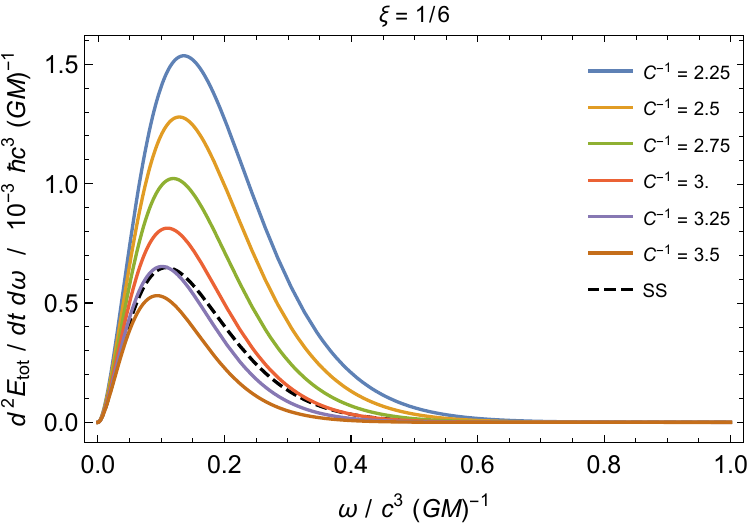}
\caption{\label{fig:SpectrumTotCompSSXi1By6} Emission spectrum of a spherically symmetric spacetime with a central object of constant density (e.g., a ``neutron star'') in equilibrium, for photon-like fields, i.e., $\xi=1/6$. We have a blackbody surface component and assume an $r$-dependent thermal distribution for the direct component.} 
\end{figure}

We can model the emission spectrum of the compact object in equilibrium by considering the two emission components separately.
For the surface emission we have blackbody radiation. The local temperature at the object surface is determined by $\der E_\text{surf}/\der t$ taking into account gravitational redshift and a graybody factor. The latter allows for the fact that for objects smaller than their unstable photon orbit at $3\Rgrav$ not all the surface emission is radiated away, but that a fraction is reflected back.
For the direct emission component, we associate each radius of emission with a Planck partition function whose average energy corresponds to the characteristic energy scale $E_\text{curv}$.
The combined spectrum is presented for the conformally coupled case in Fig.~\ref{fig:SpectrumTotCompSSXi1By6}.

\begin{figure}[!ht]
\centering
\includegraphics[width=.5\linewidth]{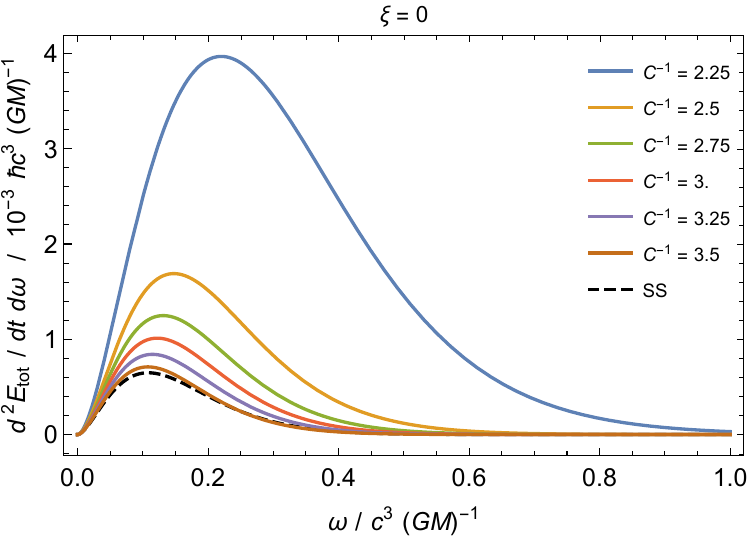}
\caption{\label{fig:SpectrumTotCompSSXi0} Emission spectrum of a spherically symmetric spacetime with a central object of constant density (e.g., a ``neutron star'') in equilibrium, for graviton-like fields, i.e., $\xi=0$. As in Fig.~\ref{fig:SpectrumTotCompSSXi1By6}, we have a blackbody surface component and assume an $r$-dependent thermal distribution for the direct component.} 
\end{figure}

Figure~\ref{fig:SpectrumTotCompSSXi0} shows the total emission spectrum for a compact object, but for the case of minimal coupling, i.e., $\xi=0$, as a representative for gravitons. As to be expected from the last term in the effective action in Eq.~\eqref{eq:Intro-EffAction}, the emission is larger as compared to the conformally coupled case due to the increased contribution from the interior of the compact object. This is most pronounced for the highest values of compactness.

The spectra of objects of higher compactness show a larger total emission and a higher peak frequency. The Schwarzschild spectrum, which due to the presence of an event horizon comprises a direct component only, lies below the emission spectra of objects with highest compactness. This underlines the relevance of the surface emission component in accordance with Fig.~\ref{fig:PowerRatioSurfToExtEscVsInvCompactness}.

Equation~\eqref{eq:EnergyfluxTotal} can be converted into an overall effective temperature, $T_\text{eff,o}$, which a static observer at infinity would associate with the total emission according to the Stefan--Boltzmann law. For compact objects with $\Rco \geq 6\,\Rgrav$, such an astronomer overestimates the temperature by less than $\sim 10\%$ when neglecting gravitational redshift,
\begin{align}
\label{eq:TEffOverall}
T_\text{eff,o}
&= \left(
\dot E_\text{tot}/4 \pi \Rco^2\, \sigma_\text{s}\right)^{1/4}\\
&=\SI{2.5e-8}{K}\, \left(\frac{f(\xi)}{0.38} \; \alpha\!\left(\frac{\Rco}{\Rgrav},\xi\right)\right)^{1/4}\, \left(\frac{M}{1.44\, M_{\odot}}\right)^{-1}\left(\frac{\Rco}{6\,\Rgrav}\right)^{-13/8}.
\end{align}
For the scalar field, 
$\sigma_\text{s}
= \pi^2 k_\text{B}^4/120 \hbar^3 c^2$ is half of the ordinary Stefan--Boltzmann constant for electromagnetic radiation. 

Next, we can estimate a characteristic timescale $\tau$ for evaporation by relating the total energy flux to the rest mass energy  
\begin{align}
\tau
&=\frac{Mc^2}{\dotEtot}\\
&=\SI{3.4e68}{yr}\; 
\left(\frac{f(\xi)}{0.38}\right)^{-1} \alpha\!\left(\frac{\Rco}{\Rgrav},\xi\right)^{-1} \left(\frac{M}{1.44\, M_{\odot}}\right)^{3}\left(\frac{\Rco}{6\,\Rgrav}\right)^{9/2}\\
\label{eq:lifetime_density}
&=\SI{3.4e68}{yr}\; 
\left(\frac{f(\xi)}{0.38}\right)^{-1} \alpha\!\left(\frac{\Rco}{\Rgrav},\xi\right)^{-1} \left(\frac{\rho}{\rhoNS}\right)^{-3/2}
\end{align}
and find that $\tau$ mainly depends on the mass density $\rho$ of the object in units of $\rhoNS = \SI{3.3e14}{g/cm^3}$, given that the correction function $\alpha$ is of order unity. The actual evaporation time depends on how the density changes when the mass decreases, 
$\rho = \rho(\Mco)$. 
For a homogeneous-density model for a neutron star, as in this paper, $\rho$ scales as $\Mco^{-2}$ in the high-density region, independent of the equation of state \cite{1973ApJ...179..277N-corrected}. This effectively resembles the accelerating evaporation behavior of a Schwarzschild black hole.

For a putative black hole with a mass as low as $\Mco=1.44\,M_\odot$, we find (in the same spirit as formula~\eqref{eq:TEffOverall} with the total flux at the radius of maximum emission, $\Rco \approx 2.5\,\Rgrav$ (see \cite{Wondrakvan-SuijlekomFalcke2023-PRL})) $T_\text{eff,o}\approx \SI{72}{nK}$ and $\tau \approx \SI{3.0e67}{yr}$, where $\tau\propto M^3$. This is a factor four shorter than a neutron star of the same mass with $\xi=0$. However, an actual stellar mass black hole of $\Mco=3\,M_\odot$ would last more than twice as long as an average neutron star. A white dwarf with $\Mco=1.3\,M_\odot$ and $\Rco=\SI{2550}{km}$ would have $T_\text{eff,o}\approx \SI{5.5}{pK}$ and a lifetime of $\tau \approx \SI{3.3e78}{yr}$ for $\xi=0$. For the supermassive black hole M87$^\ast$ with $\Mco \approx \num{6e9}\,M_\odot$, one gets $\tau \approx \SI{2e96}{yr}$.

\section{Discussion and conclusions} 
\label{sec:DiscussionConclusions}
We computed massless pair production in the gravitational field of a star or stellar remnant of constant density, based on the formalism developed in \cite{Wondrakvan-SuijlekomFalcke2023-PRL}, and found a finite energy flux that should eventually lead to their decay. This evaporation processes is similar to Hawking radiation for black holes, but does not depend on the presence of an event horizon. While an observer at infinity only measures the direct emission component in the case of black hole evaporation, in the case of a horizonless compact object she also measures a surface emission component due to absorbed particles, which can be significant. Effective temperature, energy flux, and timescale of the evaporation depend on mass $\Mco$ and compactness $C$ (i.e., inverse dimensionless radius) of the object.

Despite the fact that the emission is generated potentially far from the central object, causality is not violated, neither for the Hawking effect nor for gravitational curvature radiation, as a radially extended source geometry is consistent with the Heisenberg uncertainty relation. For gravitational curvature radiation,
virtual particles produced at $r$ have an energy for a local observer amounting to 
$\Delta E=2^{3/4} \hbar c/5^{1/4} \Rgrav \times (r/\Rgrav)^{-3/2}$
(see Sec.~\ref{sec:GravPtcleProd}).
Being quantum fluctuations, where $\Delta E$ is of this energy scale, their time of existence has an upper bound implied by the uncertainty relation
$\Delta t \lesssim \hbar/2 \Delta E$.
This time needs to be sufficient to reach the event horizon or compact object’s surface, i.e., to travel for 
$\Delta r = r - \Rco$, 
for ensuring positive-energy observables. Inserting the speed of light, this leads to the inequality
$\Delta r/\Rco (1 + \Delta r/\Rco)^{3/2} \times C^{1/2} \lesssim 5^{1/4} / 2^{7/4}$
which holds true for all values of compactness $C = \Rgrav/\Rco \leq 1/2$, i.e., even for the most compact matter accumulations including black holes. 

The emission process itself raises other interesting questions, for example
about describing the decay process at the level of individual atomic nuclei. For astrophysical objects, the rest mass of a baryon lies well above the amount of energy lost by a single pair. Hence, whether an isolated nucleon or electron is able to decay via this mechanism is unclear. However, at lower energy scales, compact objects offer sea quarks and phonons to interact with. The accumulated energy loss via such channels could then eventually lead to the decay of a single proton. 
Conserving electric charge and $B-L$, the difference between baryon and lepton number, the proton could have a decay channel into a pion and a positron, which then would annihilate with an electron. This needs to happen statistically, analogously to radioactive decay.%
\footnote{%
The decay equation induced by the total energy flux,
$\der (M c^2)/\der t 
= -\der E_\text{tot}/\der t 
\propto -M\, \rho^{3/2}$,
resembles the radioactive decay law 
in the case that the density is independent of the mass such as for bodies with negligible self-gravitation.}

Our approach involves two idealizations of the stationary astrophysical setting. First, we assume an optically thick compact object which implies a complete absorption of infalling radiation and a purely thermal emission from the surface. 
A more refined treatment of absorption and energy transport would involve a frequency-dependent spectral optical depth and generalizations beyond the geometric optics limit.
Second, realistic stellar remnants do not follow a constant density profile but rather a density distribution $\rho(r)\propto 1 - (r/\Rco)^2$ \cite{Tolman1939a,LattimerPrakash2001a}. 
This should introduce corrections of order unity. 
Moreover, our approach uses an approximation which is second order in spacetime curvature and valid to arbitrary, but finite order in proper time. While it does recover the Hawking and Schwinger effects, there might be additional impact from re-summing the infinitely many terms which are inaccessible with current methods. 
Finally, like Hawking radiation, this effect is not experimentally verified and there is little hope that this can ever be achieved for macroscopic objects -- apart from experiments in analog gravity. 

Barring these caveats, a key theoretical result is that the characteristic evaporation time scales with the average density of the star as $\tau\propto\rho^{-3/2}$. For this reason stellar mass black holes and neutron stars have rather comparable lifetimes ($\tau \sim 10^{67-68}\,$yr), while white dwarfs can survive much longer ($\tau \gtrsim \SI{e78}{yr}$), eventually followed by supermassive black holes ($\tau \gtrsim \SI{e96}{yr}$). 

\begin{figure}[th]
\centering
\includegraphics[width=.7\linewidth]{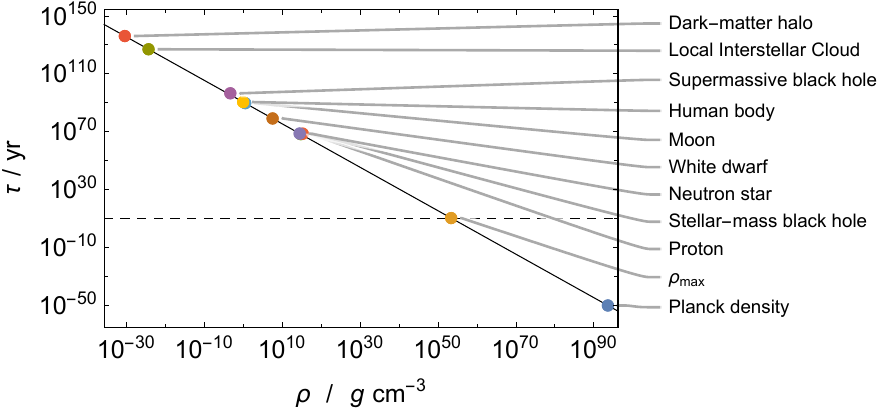}
\caption{\label{fig:LifetimeVsMassDensity} Characteristic timescale for evaporation for several objects discussed in the text as a function of mass density. The black solid line corresponds to Eq.~\eqref{eq:lifetime_density}, the dashed line to the age of the universe.}
\end{figure}

In principle, the process could also be applicable to other astrophysical objects: 
The Moon ($\rho \approx \SI{3.4}{g/cm^3}$) has $\tau \sim \SI{3e89}{yr}$, a body with the density of water has $\tau \sim \SI{e90}{yr}$, the Local Interstellar Cloud ($\rho \approx \SI{5e-25}{g/cm^3}$) has $\tau \sim \SI{e127}{yr}$, and a dark matter halo of a supercluster (mass of $10^{17}\,M_{\odot}$, size of $\SI{160}{Mpc}$) has $\tau \sim \SI{e135}{yr}$.
For the Moon this translates to the decay of one proton roughly every $\sim\SI{e40}{yr}$, which, like Hawking radiation, is not directly detectable. 
These lifetimes are put into perspective in Fig.~\ref{fig:LifetimeVsMassDensity}.
Of course, these examples ignore other astrophysical evolution and decay channels and the induced change in mass density. Therefore one should consider these timescales only as absolute theoretical upper limits showing at least that the presence of this effect is not ruled out by the existence of astrophysical objects.

While the highest known mass densities occur in the interior of neutron stars, Eq.~\eqref{eq:lifetime_density} predicts the existence of a highest quasi-stable density scale for the present age of the universe. It is
$\rho_\text{max} 
\approx \SI{3e53}{g/cm^3} 
\sim \num{e39}\, \rhoNS 
\sim \num{e-40}\, \rho_\text{Pl}$, which lies much below the typical quantum-gravity scale given by the Planck density 
$\rho_\text{Pl} \approx \SI{5e93}{g/cm^3}$.
This indicates that stable objects at Planckian scales  as predicted by string theory (see, e.g., \cite{Nicolini:2019irw}) should be absent because they had a characteristic lifetime of few Planck times, 
$\tau 
\approx 4\,t_\text{Pl} 
\sim \SI{e-43}{s}$.

For a neutron star, the evaporation process can continue only until their minimum mass ($\sim 0.1\,M_\odot$ \cite{HarrisonThorneWakano1965a-updated}) is reached, when it will explode and produce an observable burst of high-energy particles and neutrinos  \cite{BlinnikovImshennikNadezhin1990a,YudinDunina-BarkovskayaBlinnikov2023a-updated}. Given the long timescales, we do not expect any neutron stars formed in our current universe to undergo such an evolution. One could speculate that fossil neutron stars from a previous universe might still be around and be near the critical mass. This is only possible if inflation does not prevent universes from occupying the same phase space. If present, fossil neutron stars would now be growing by accretion from the intergalactic medium and the cosmic microwave background rather than shrinking, unless some instability would make them undergo such a phase transition after all.

Star formation rates in hypothetical counterfactual universes have been calculated by varying different cosmological parameters~\cite{BoussoLeichenauer2010a,BarnesElahiSalcido2018a,OhPeacockKhochfar2022a}. This is used to estimate the likelihood of the current set of these parameters in the landscape of possible multiverses. Non-detection of such an isolated neutron star or white dwarf population, e.g., via microlensing \cite{MrozWyrzykowski2021a,MrozUdalskiWyrzykowski2021a}, might be used to set constraints on the recurrence of star forming universes in a particular multiverse scenario. However, at present, this remains speculative at best and the likelihood of detecting such objects is presumably small. 

The tidal-forces description of gravitational particle production (see, e.g., \cite[pp.~367--368]{ShapiroTeukolsky1983a}, \cite[p.~264]{BirrellDavies1984a}, and \cite{Dey:2017yez,Ong:2020hti}) naturally shows that particles are created outside the object. 
For energy conservation, one might expect the presence of negative energy modes such as violations of the classical energy conditions are encountered already in flat spacetimes, e.g., in the context of the Casimir effect~\cite{Klimchitskaya:2009cw}.
In contrast, one might rather think in terms of interactions of positive energy modes where, e.g., phonons couple to gravitons which undergo pair creation.
The total energy is conserved when considering the energy of the gravity and matter system combined (such as in Ref.~\cite{Candelas:1980zt,Parikh:1999mf}), in the same spirit as combining the entropies of causal horizons and of matter in the generalized second law of thermodynamics.

For the future, it will be interesting to think about this evaporation process in the light of the information paradox \cite{Maldacena2020a}. Given that the emission of virtual pairs is separated from the location of decaying matter within the limits of the Heisenberg uncertainty principle, and that it is not a priori clear which of the two particles escapes or is absorbed by the surface, it is not immediately obvious how quantum information can be preserved within the context of gravitational pair creation. 
Further work is needed to address these fundamental questions.

\acknowledgments
We thank the referee for comments which helped to improve and clarify the manuscript.
This work was supported by 
the ERC Synergy Grant ``BlackHolistic'' 
(HF, R.\ Fender, and S.\ Markoff), 
the NWO Spinoza Prize (HF),
a grant from NWO NWA 6201348 (HF and WDvS),
and the Excellence Fellowship from Radboud University (MFW). 
HF acknowledges useful discussions with Ethan Siegel about multiverses on Mastodon.

\bibliographystyle{JHEP} 

\begin{thebibliography}{10}

\bibitem{AdamsLaughlin1997a}
F.C.~{Adams} and G.~{Laughlin}, \emph{{A dying universe: the long-term fate and
  evolution of astrophysical objects}},
  {\emph{Rev.\ Mod.\ Phys.} {\bfseries 69} (1997) 337}
  [\href{https://arxiv.org/abs/astro-ph/9701131}{{\ttfamily
  astro-ph/9701131}}].

\bibitem{Hawking1974a}
S.W.~{Hawking}, \emph{{Black hole explosions?}},
  {\emph{\nat} {\bfseries 248} (1974) 30}.

\bibitem{Hawking1975a}
S.W.~{Hawking}, \emph{{Particle creation by black holes}},
  {\emph{Commun.\ Math.\ Phys.} {\bfseries 43} (1975) 199}.

\bibitem{NathFileviez-Perez2007a}
P.~{Nath} and P.~{Fileviez P{\'e}rez}, \emph{{Proton stability in grand unified
  theories, in strings and in branes}},
  {\emph{\physrep}
  {\bfseries 441} (2007) 191}
  [\href{https://arxiv.org/abs/hep-ph/0601023}{{\ttfamily hep-ph/0601023}}].

\bibitem{Caplan2020a}
M.E.~{Caplan}, \emph{{Black dwarf supernova in the far future}},
  {\emph{\mnras} {\bfseries 497}
  (2020) 4357} [\href{https://arxiv.org/abs/2008.02296}{{\ttfamily
  2008.02296}}].

\bibitem{Zeldovich1976a}
Y.B.~{Zeldovich}, \emph{{A new type of radioactive decay: gravitational
  annihilation of baryons}},
  {\emph{Phys.\ Lett.\ A}
  {\bfseries 59} (1976) 254}.

\bibitem{Wondrakvan-SuijlekomFalcke2023-PRL}
M.F.~{Wondrak}, W.D.~{van Suijlekom} and H.~{Falcke}, \emph{{Gravitational Pair
  Production and Black Hole Evaporation}},
  {\emph{Phys.\ Rev.\ Lett.}
  {\bfseries 130} (2023) 221502}
  [\href{https://arxiv.org/abs/2305.18521}{{\ttfamily 2305.18521}}].

\bibitem{MathurMehta2023a}
S.D.~{Mathur} and M.~{Mehta}, \emph{{The universality of black hole
  thermodynamics}},
  {\emph{Int.\ J.\ Mod.\ Phys.\ D} {\bfseries 32} (2023) 2341003}
  [\href{https://arxiv.org/abs/2305.12003}{{\ttfamily 2305.12003}}].

\bibitem{Mathur:2024mvo}
S.D.~Mathur and M.~Mehta, \emph{{The universal thermodynamic properties of
  extremely compact objects}}, {\emph{Class. Quant. Grav.} {\bfseries 41}
  (2024) 235011} [\href{https://arxiv.org/abs/2402.13166}{{\ttfamily
  2402.13166}}].

\bibitem{ShapiroTeukolsky1983a}
S.L.~{Shapiro} and S.A.~{Teukolsky}, \emph{{Black Holes, White Dwarfs,
and Neutron Stars. The Physics of Compact Objects}}, WILEY-
VCH Verlag GmbH \& Co. KGaA, Weinheim (2004).

\bibitem{BirrellDavies1984a}
N.D.~{Birrell} and P.C.W.~{Davies}, \emph{{Quantum Fields in Curved Space}}, Cambridge Monographs on Mathematical Physics, 
  Cambridge University Press, Cambridge, England (1984).

\bibitem{Dey:2017yez}
R.~Dey, S.~Liberati and D.~Pranzetti, \emph{{The black hole quantum
  atmosphere}}, {\emph{Phys. Lett. B} {\bfseries 774} (2017) 308}
  [\href{https://arxiv.org/abs/1701.06161}{{\ttfamily 1701.06161}}].

\bibitem{Ong:2020hti}
Y.C.~Ong and M.R.R.~Good, \emph{{Quantum atmosphere of Reissner-Nordstr\"om
  black holes}}, {\emph{Phys. Rev. Res.} {\bfseries 2} (2020) 033322}
  [\href{https://arxiv.org/abs/2003.10429}{{\ttfamily 2003.10429}}].

\bibitem{Parker:1968mv}
L.~Parker, \emph{{Particle creation in expanding universes}}, {\emph{Phys. Rev.
  Lett.} {\bfseries 21} (1968) 562}.

\bibitem{ZeldovichStarobinskii1972a-updated}
Y.B.~{Zel'dovich} and A.A.~{Starobinski{\v{i}}}, \emph{{Particle Production and
  Vacuum Polarization in an Anisotropic Gravitational Field}}, 
  {\emph{Zh. Eksp. Teor. Fiz.} {\bfseries 61} (1971) 2161} [{\emph{Sov. Phys. JETP} {\bfseries 34} (1972) 1159]}.

\bibitem{Visser:2001kq}
M.~Visser, \emph{{Essential and inessential features of Hawking radiation}},
  {\emph{Int. J. Mod. Phys. D} {\bfseries 12} (2003) 649}
  [\href{https://arxiv.org/abs/hep-th/0106111}{{\ttfamily hep-th/0106111}}].

\bibitem{HossenfelderSchwarzGreiner2003a}
S.~{Hossenfelder}, D.J.~{Schwarz} and W.~{Greiner}, \emph{{Particle production
  in time-dependent gravitational fields: the expanding mass shell}},
  {\emph{Class. Quant. Grav.} {\bfseries 20} (2003) 2337}
  [\href{https://arxiv.org/abs/gr-qc/0210110}{{\ttfamily gr-qc/0210110}}].

\bibitem{Vachaspati:2006ki}
T.~Vachaspati, D.~Stojkovic and L.M.~Krauss, \emph{{Observation of incipient
  black holes and the information loss problem}},
  {\emph{Phys. Rev. D}
  {\bfseries 76} (2007) 024005}
  [\href{https://arxiv.org/abs/gr-qc/0609024}{{\ttfamily gr-qc/0609024}}].

\bibitem{Unruh1976a}
W.G.~{Unruh}, \emph{{Notes on black-hole evaporation}},
  {\emph{\prd} {\bfseries 14}
  (1976) 870}.

\bibitem{Lima:2018ifz}
C.A.U.~Lima, F.~Brito, J.A.~Hoyos and D.A.T.~Vanzella, \emph{{Probing the Unruh
  effect with an accelerated extended system}}, {\emph{Nature Commun.}
  {\bfseries 10} (2019) 3030}
  [\href{https://arxiv.org/abs/1805.00168}{{\ttfamily 1805.00168}}].

\bibitem{Schwinger1951a}
J.~{Schwinger}, \emph{{On Gauge Invariance and Vacuum Polarization}},
  {\emph{Phys. Rev.}
  {\bfseries 82} (1951) 664}.

\bibitem{BarvinskyVilkovisky1990a}
A.O.~{Barvinsky} and G.A.~{Vilkovisky}, \emph{{Covariant perturbation theory
  (II). Second order in the curvature. General algorithms}},
  {\emph{Nucl. Phys. B}
  {\bfseries 333} (1990) 471}.

\bibitem{Buchdahl1959a}
H.A.~{Buchdahl}, \emph{{General Relativistic Fluid Spheres}},
  {\emph{Phys. Rev.}
  {\bfseries 116} (1959) 1027}.

\bibitem{Wondrakvan-SuijlekomFalcke2023-Reply}
M.F.~Wondrak, W.D.~van Suijlekom and H.~Falcke, \emph{{Reply to ''Comment on
  'Gravitational Pair Production and Black Hole Evaporation'{}''}},
  {\emph{Phys. Rev. Lett.} {\bfseries 133} (2024) 229002}
  [\href{https://arxiv.org/abs/2308.12326}{{\ttfamily 2308.12326}}].

\bibitem{FerreiroNavarro-SalasPla2023a-Comment}
A.~Ferreiro, J.~Navarro-Salas and S.~Pla, \emph{{Comment on
  \textquotedblleft{}Gravitational Pair Production and Black Hole
  Evaporation\textquotedblright{}}}, {\emph{Phys. Rev. Lett.} {\bfseries 133}
  (2024) 229001} [\href{https://arxiv.org/abs/2306.07628}{{\ttfamily
  2306.07628}}].

\bibitem{CodelloZanusso2013a}
A.~{Codello} and O.~{Zanusso}, \emph{{On the non-local heat kernel expansion}},
  {\emph{J. Math. Phys.} {\bfseries 54} (2013) 013513}
  [\href{https://arxiv.org/abs/1203.2034}{{\ttfamily 1203.2034}}].

\bibitem{El-Menoufi2016a}
B.K.~El-Menoufi, \emph{Quantum gravity of {Kerr}-{Schild} spacetimes and the
  logarithmic correction to {Schwarzschild} black hole entropy},
  {\emph{J. High Energy Phys.} {\bfseries 2016} (2016) 35}.

\bibitem{Chernodub2023a}
M.N.~{Chernodub}, \emph{{Conformal anomaly and gravitational pair production}},
  {\emph{arXiv e-prints} (2023)
  arXiv:2306.03892} [\href{https://arxiv.org/abs/2306.03892}{{\ttfamily
  2306.03892}}].

\bibitem{Schwarzschild1916b-adapted}
K.~{Schwarzschild}, \emph{{{\"U}ber das Gravitationsfeld einer Kugel aus
  inkompressibler Fl{\"u}ssigkeit nach der Einsteinschen Theorie}},
  {\emph{Sitzungsberichte der K{\"o}niglich Preu\ss{}ischen Akademie der
  Wissenschaften zu Berlin} {\bfseries 1916.XVIII} (1916) 424}.

\bibitem{Tolman1939a}
R.C.~{Tolman}, \emph{{Static Solutions of Einstein's Field Equations for
  Spheres of Fluid}},
  {\emph{Phys. Rev.}
  {\bfseries 55} (1939) 364}.

\bibitem{OppenheimerVolkoff1939a}
J.R.~{Oppenheimer} and G.M.~{Volkoff}, \emph{{On Massive Neutron Cores}},
  {\emph{Phys. Rev.}
  {\bfseries 55} (1939) 374}.

\bibitem{CardosoPani2019a}
V.~{Cardoso} and P.~{Pani}, \emph{{Testing the nature of dark compact objects:
  a status report}},
  {\emph{Living Rev. Relativ.} {\bfseries 22} (2019) 4}
  [\href{https://arxiv.org/abs/1904.05363}{{\ttfamily 1904.05363}}].

\bibitem{Synge1966a}
J.L.~{Synge}, \emph{{The escape of photons from gravitationally intense
  stars}}, {\emph{\mnras}
  {\bfseries 131} (1966) 463}.

\bibitem{1973ApJ...179..277N-corrected}
M.~{Nauenberg} and G.~{Chapline}, Jr., \emph{{Determination of Properties of
  Cold Stars in General Relativity by a Variational Method}},
  {\emph{\apj} {\bfseries 179} (1973)
  277}.

\bibitem{LattimerPrakash2001a}
J.M.~{Lattimer} and M.~{Prakash}, \emph{{Neutron Star Structure and the
  Equation of State}}, {\emph{\apj}
  {\bfseries 550} (2001) 426}
  [\href{https://arxiv.org/abs/astro-ph/0002232}{{\ttfamily
  astro-ph/0002232}}].

\bibitem{Nicolini:2019irw}
P.~Nicolini, E.~Spallucci and M.F.~Wondrak, \emph{{Quantum Corrected Black
  Holes from String T-Duality}},
  {\emph{Phys. Lett. B}
  {\bfseries 797} (2019) 134888}
  [\href{https://arxiv.org/abs/1902.11242}{{\ttfamily 1902.11242}}].

\bibitem{HarrisonThorneWakano1965a-updated}
B.K.~{Harrison}, K.S.~{Thorne}, M.~{Wakano} and J.A.~{Wheeler},
  \emph{{Gravitation Theory and Gravitational Collapse}}, University of Chicago
  Press, Chicago (1965).

\bibitem{BlinnikovImshennikNadezhin1990a}
S.I.~{Blinnikov}, V.S.~{Imshennik}, D.K.~{Nadezhin}, I.D.~{Novikov},
  T.V.~{Perevodchikova} and A.G.~{Polnarev}, \emph{{Explosion of a Low-Mass
  Neutron Star}}, {\emph{\sovast} {\bfseries 34} (1990) 595}.

\bibitem{YudinDunina-BarkovskayaBlinnikov2023a-updated}
A.V.~{Yudin}, N.V.~{Dunina-Barkovskaya} and S.I.~{Blinnikov}, \emph{{Thermal
  Neutrinos from the Explosion of a Minimum-Mass Neutron Star}},
  {\emph{Astron. Lett.}
  {\bfseries 48} (2022) 497}
  [\href{https://arxiv.org/abs/2301.10003}{{\ttfamily 2301.10003}}].

\bibitem{BoussoLeichenauer2010a}
R.~{Bousso} and S.~{Leichenauer}, \emph{{Predictions from star formation in the
  multiverse}}, {\emph{\prd}
  {\bfseries 81} (2010) 063524}
  [\href{https://arxiv.org/abs/0907.4917}{{\ttfamily 0907.4917}}].

\bibitem{BarnesElahiSalcido2018a}
L.A.~{Barnes}, P.J.~{Elahi}, J.~{Salcido}, R.G.~{Bower}, G.F.~{Lewis},
  T.~{Theuns} et~al., \emph{{Galaxy formation efficiency and the multiverse
  explanation of the cosmological constant with EAGLE simulations}},
  {\emph{\mnras} {\bfseries 477}
  (2018) 3727} [\href{https://arxiv.org/abs/1801.08781}{{\ttfamily
  1801.08781}}].

\bibitem{OhPeacockKhochfar2022a}
B.K.~{Oh}, J.A.~{Peacock}, S.~{Khochfar} and B.D.~{Smith}, \emph{{The fate of
  baryons in counterfactual universes}},
  {\emph{\mnras} {\bfseries 517}
  (2022) 59}.

\bibitem{MrozWyrzykowski2021a}
P.~{Mr{\'o}z} and {\L}.~{Wyrzykowski}, \emph{{Measuring the Mass Function of
  Isolated Stellar Remnants with Gravitational Microlensing I. Revisiting the
  OGLE-III Dark Lens Candidates}},
  {\emph{\actaa} {\bfseries 71}
  (2021) 89} [\href{https://arxiv.org/abs/2107.13701}{{\ttfamily 2107.13701}}].

\bibitem{MrozUdalskiWyrzykowski2021a}
P.~{Mroz}, A.~{Udalski}, L.~{Wyrzykowski}, J.~{Skowron}, R.~{Poleski},
  M.~{Szymanski} et~al., \emph{{Measuring the mass function of isolated stellar
  remnants with gravitational microlensing. II. Analysis of the OGLE-III
  data}}, {\emph{arXiv
  e-prints} (2021) arXiv:2107.13697}
  [\href{https://arxiv.org/abs/2107.13697}{{\ttfamily 2107.13697}}].

\bibitem{Klimchitskaya:2009cw}
G.L.~Klimchitskaya, U.~Mohideen and V.M.~Mostepanenko, \emph{{The Casimir force
  between real materials: Experiment and theory}},
  {\emph{Rev. Mod. Phys.}
  {\bfseries 81} (2009) 1827}
  [\href{https://arxiv.org/abs/0902.4022}{{\ttfamily 0902.4022}}].

\bibitem{Candelas:1980zt}
P.~Candelas, \emph{{Vacuum Polarization in Schwarzschild Space-Time}},
  {\emph{Phys. Rev. D} {\bfseries 21} (1980) 2185}.

\bibitem{Parikh:1999mf}
M.K.~Parikh and F.~Wilczek, \emph{{Hawking radiation as tunneling}},
  {\emph{Phys. Rev. Lett.} {\bfseries 85} (2000) 5042}
  [\href{https://arxiv.org/abs/hep-th/9907001}{{\ttfamily hep-th/9907001}}].

\bibitem{Maldacena2020a}
J.~Maldacena, \emph{Black holes and quantum information},
  {\emph{Nat. Rev. Phys.} {\bfseries 2} (2020) 123}.

\end{thebibliography}

\providecommand{\href}[2]{#2}\begingroup\raggedright\endgroup

\end{document}